# Evolutionary relations between different types of Magnetized Compact Objects

Vladimir Lipunov*,[a,b], Valeria Grinshpun[a], Daniil Vlasenko[a,b],

[a] Lomonosov Moscow State University, Physics Department, Russia GSP-1, Leninskie Gory, Moscow, 119991, Russia
[b] Lomonosov Moscow State University, Sternberg Astronomical Institute, Universitetskiy pr. 13, Moscow 119234, Russia

ABSTRACT

The numerous compact sources associated with neutron stars and white dwarfs discovered in recent decades are analyzed in terms of the Gravimagnetic Rotator model (GMR paradigm - Lipunov, 1987a; Lipunov, 1992). We offer the instrument for understanding of various observed features and evolutionary relationships of neutron stars and white dwarfs. We depict in a single diagram all objects from radio pulsars and dwarf novae to ultra luminous X-ray sources and a radio pulsating white dwarf. This diagram directly demonstrates the genetic link between different types of compact sources thereby making it possible to confirm and illustrate clearly the established evolutionary connections–such as that between bulge X-ray sources and millisecond pulsars. This approach allows us to understand the evolutionary status of Ultra Luminous X-ray sources. In addition, we propose an additional evolutionary branch of the formation of Magnetars. When our work was completed, an article by Kirsten et al.2021, was published, which reports the localization of FRB 20200120 in one of the globular clusters of the galaxy M81. This shows that the accretion-induced collapse scenario of the white dwarf (Lipunov & Postnov, 1985), considered in detail in this work, is a possible genealogical branch of Magnetar production.

## 1. Introduction

There are many types of compact sources of electromagnetic radiations, connected with neutron stars and white dwarfs, that were discovered during last decades. We will consider them within the model of the gravitationally magnetic rotator (Lipunov, 1987a; Lipunov, 1992).

The general laws of normal binary stars evolution were formulated by Crawford (1955), Snezhko (1967), van den Heuvel & Heise (1972); Tutukov et al. (1973); van den Heuvel (1977).

Kardashev (1964) and Pacini (1967) were the first to consider the idea that the model of a magnetic rotator can explain the energy sources of compact astrophysical objects. This idea became especially popular immediately after the discovery of radio pulsars (Hewish et al. 1968) and their interpretation as magnetized neutron stars with dipole magnetic fields (Gold, 1969). On the other hand, Zeldovich &Novikov (1966) and Shklovsky (1967) paid special attention to the accretion mechanism (Zeldovich, 1964 ; Salpeter, 1964) and the powerful gravitational field of magnetized neutron stars. In fact, even before the discovery of radio and X-ray (Giacconi et al.,1972) pulsars, young graduate students of Ya.B. Zeldovich with co-authors (Amnuel et al.,1968; Bisnovaty-Kogan & Friedman 1969, Zeldovich & Shakura 1969, Shvartsman 1972) actually predicted the phenomenon of pulsar type of X-ray emission of accreting neutron stars.

The next 10 years were marked by the discovery of dozens and hundreds of objects that sometimes had very special astrophysical properties while also being associated with compact stars — neutron stars and white dwarfs: radio pulsars, X-ray pulsars, X-ray bursters, cataclysmic variable stars, polars, intermediate polars, millisecond radio pulsars, X-ray Novae, and SS433-type sources.

As a result, it became important to study a unified approach of such various objects if possible. In this regard, a generalized model of a gravi-magnetic rotator (GMR Lipunov, 1982a; 1987a; see a monograph Lipunov 1992) has emerged which may unite a whole variety of compact sources on energy release mechanisms and link them within the framework of spin evolution .

---

* corresponding author:   lipunov2007@gmail.com

## 2. The Classification of GMR

The gravimagnetic rotator (GMR) is an object with a mass M, magnetic dipole moment *μ*, angular momentum **K = I ω**, where *ω = 2π / p* is the circular frequency of rotation and *p* is the spin period (Fig. 1).

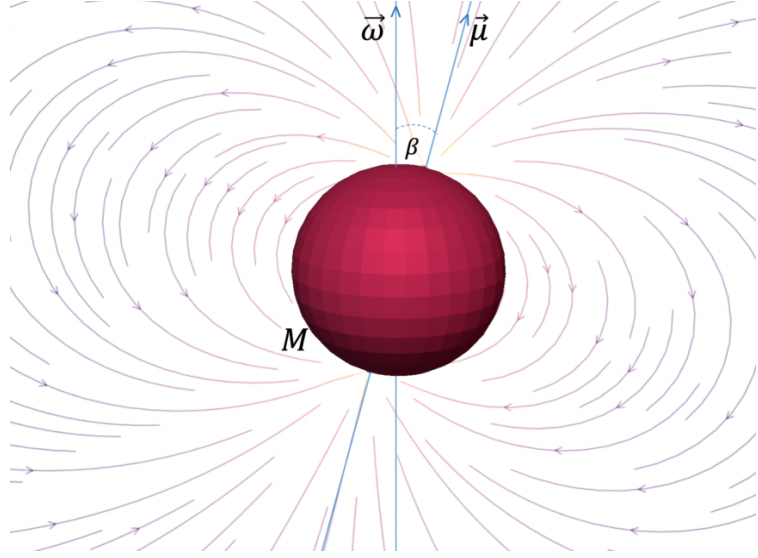

**Fig. 1.** Gravimagnetic rotator. M, **ω**, **μ** are the mass, angular velocity of rotation, and magnetic dipole moment, respectively.

It is clear that the solution of the problem of gravitational and electromagnetic interaction of such an object with the surrounding plasma capable of accretion is a complex 3D magneto-hydrodynamic problem in the gravitational field of the rotator. As a zero approximation Lipunov (1987a) proposed a qualitative approach to isolate the main modes of interaction based on the relationship between the main characteristic radius of interaction between the rotator and the environment.

The problem involves four characteristic radii responsible for the gravitational and electromagnetic parts: $R_{stop}$ is the stop radius; $R_c = (GM/\omega^2)^{1/3}$ is the corotation radius; $R_l = c/\omega$ is the light cylinder radius, and $R_G = 2GM/v_\infty^2$ is the gravitational capture radius (the Bondi-Hoyle radius). In binary systems, an additional characteristic dimension *a* - the distance between the components - appears. In these formulas, $v_\infty$ is the speed of the rotator relative to the environment. For simplicity, we assume that the motion of the rotator is supersonic (as is usually the case under astrophysical conditions).

The stop radius is the characteristic distance at which the outward pressure of the electromagnetic forces $P_m$ and the accretion pressure $P_a$ of the matter falling inwards are balanced:

$$P_m = P_a \qquad (1)$$

If balance (1) is established outside the radius of the light cylinder $R_l$, then the accretion of the surrounding plasma is stopped by the relativistic wind of a compact star, as was first considered by Shwartzman (1970):

$$P_m = L_m/4\pi R^2 c \ , \quad R_{stop} > R_l \qquad (2)$$

where $L_m$ is the total power of the compact star in the relativistic wind. Of course, we mean approximate estimates, in which we neglect the angular dependences.

If the plasma penetrates under the light cylinder, then it is stopped by the pressure of a quasi-static magnetic field:

$$P_m = B^2/8\pi , \quad R_{stop} < R_l \qquad (3)$$

and the stopping radius is usually called the Alfven radius (Davidson and Ostriker, 1973):

$$R_{stop} = \begin{cases} R_{sh} - Shwartzman\ radius\ R_{stop} > R_l \\ R_A - Alfven\ radius\ if\ R_{stop} < R_l \end{cases} \qquad (4)$$

In the classification (see also Table 1), we used an additional parameter, namely the accretion rate $\dot{M}$. The fact is that the astrophysical properties of a compact star vary significantly if the accretion rate exceeds the critical limit $\dot{M}_{cr}$ corresponding to Eddington limit (Lipunov, 1982b):

$$\dot{M}_{cr}\ GM/R_{stop} = L_{Edd} = 4\pi GMc/k_T \qquad (5)$$

$$\dot{M}_{cr} = 4\pi c\ R_{stop} k_T \approx 10^{-6}\ R_8\ M_\odot/yr \qquad (6)$$

where $R_8 = R_{stop}/10^8 cm$ – is the Alfven radius in units of $10^8$ cm, and $k_T$ is the Thompson cross section for photon scattering on free electrons calculated per unit mass.

The maximal luminosity of the neutron star accretor (under the critical accretion regime) is:

$$L_{max} = \dot{M}_{cr}\ GM/R_x \approx 1.3\ 10^{40}\ R_8/R_6\ [erg/s] \qquad (7)$$

where $R_6 = R_x/10^6 cm$ - star radius.

Supercritical accretion rates occur in massive binary systems when a normal star overflows its Roche lobe. However, this accretion mode is always accompanied with the formation of an accretion disk. Moreover, if $\dot{M} > \dot{M}_{cr}$ radiation pressure begins to exceed gravitational pressure and the excess matter is ejected from the disk surface in the form of a kind of stellar wind (Shakura & Sumyaev, 1973; Lipunov, 1982b). Thus, we have a limited set of possible modes of interaction of the gravimagnetic rotator with the accreting plasma. Let us emphasize one very significant, but often misunderstood circumstance. The energy release in the latter condition of the definition of the super-critical regime is determined at the stop radius and not at the surface of the accreting star. We do not consider supercritical regimes for white dwarfs because they have large radii and are mainly observed in low-mass binary systems.

Table 1. Classification of neutron stars

| Abbreviation | Type | Characteristic Radius Relation | Accretion Rate | Observational Appearances (Lipunov, 1992a) |
|---|---|---|---|---|
| **E** | Ejector | $R_{stop} > R_G$ <br> $R_{stop} > R_l$ | $\dot{M}_c \leq \dot{M}_{cr}$ | Radio pulsars, Millisecond pulsars, SGR, AXPs, FRB Repeaters |
| **P** | Propeller | $R_c < R_{stop}$ <br> $R_{stop} \leq R_G$ <br> $R_{stop} \leq R_l$ | $\dot{M}_c \leq \dot{M}_{cr}$ | X-ray transients? Rapid burster? |
| **A** | Accretor | $R_{stop} \leq R_G$ <br> $R_{stop} \leq R_l$ | $\dot{M}_c \leq \dot{M}_{cr}$ | X-ray pulsars, X-ray Bursters, ULX, |

|   |   |   |   | Cataclysmic variables, Intermediate polars |
|---|---|---|---|---|
| **G** | Georotator | $R_G < R_{st}$ <br> $R_{stop} \leq R_c$ | $\dot{M}_c \leq \dot{M}_{cr}$ | ? |
| **M** | Magnetor | $R_{stop} > a$ | $\dot{M}_c \leq \dot{M}_{cr}$ | AM Her type (Polars) |
| **SE** | Super-Ejector | $R_{stop} > R_l$ | $\dot{M}_c > \dot{M}_{cr}$ | ? |
| **SP** | Super-Propeller | $R_c < R_{st}$ <br> $R_{stop} \leq R_l$ | $\dot{M}_c > \dot{M}_{cr}$ | ? |
| **SA** | Super-Accretor | $R_{stop} \leq R_c$ <br> $R_{stop} \leq R_G$ | $\dot{M}_c > \dot{M}_{cr}$ | SS433? <br> Ultra soft X-sources? |

The stage of GMR is most sensitive to the gravimagnetic parameter (cf Davies & Pringle 1981, Lipunov 1987a):

$$y = \dot{M}_c/\mu^2 \qquad (8)$$

Here, $\dot{M}_c$ is the potential accretion rate on the GMR as defined by the Bondi-Hoyle formula $\dot{M}_c \equiv \pi R_G^2 \rho_\infty v_\infty$ (Bondi and Hoyle, 1944; Hoyle et al.,1948; Bondi, 1952), $\rho_\infty$ is the density of the background plasma and $v_\infty$ is the relative velocity of the GMR. In principle, the *y* parameter may also incorporate the velocity of motion: the $y = \dot{M}_c v_\infty/\mu^2$ (Osmikin & Prokhorov, Personal communication, see Lipunov, 1992). However, in this paper we adopt the initial form of the gravimagnetic parameter and point out that all the ensuing uncertainties should depend linearly on the relative velocity, which we set equal to $v_\infty = 100$ km/s for all GMR. The *y*-parameter varies over 20 orders of magnitude in astrophysical reality. Hence the uncertainty of the velocity, which can reach several hundred km/s, has little effect on our picture and no effect on our conclusions.

The *p-y* diagrams were presented 30 years ago (Lipunov 1987a) and we will check how relevant they are at present. From the point of view of the GMR model, there is no fundamental (principal) difference between neutron stars and white dwarfs. In the case of disk accretion, both of them are "pressed" to the so-called equilibrium line, which corresponds to the condition $R_A \approx R_c$.

The *p-y* diagrams are convenient for representing the evolutionary relationship between different types of Gravimagnetic Rotators.

These diagrams were presented several decades ago and in this report we will show how do they look at the current level of observations in 2021.

## 3. Universal *p-y* diagram after 30 years.

We will demonstrate the versatility of the model of a GMR in the light of discoveries made over the past three decades.

Fig. 2 presents the observational parameters of 8 types of astrophysical sources in the model of a gravity-magnetic rotator. We make one important remark. In this diagram we don't show supercritical regimes and the Georotator mode. Firstly, the position of these types of a Georotator is dependent on an additional parameter - the magnetic dipole moment. Secondly, we do not clutter the diagram with supercritical modes of interaction of a gravitating magnetized star with the surrounding plasma because concrete observational candidates are listed here (see, however, Erkut, Hakan 2017, which proposes a superpropeller model for explaining non-pulsating ULX sources). For those reasons, we do not show the Georotator stage (G). We emphasize that the **p-y** diagram is the only diagram that simultaneously shows not only various astrophysical objects associated with one type of compact star (for example, a neutron star), but also various types of compact stars — white dwarfs and neutron stars.

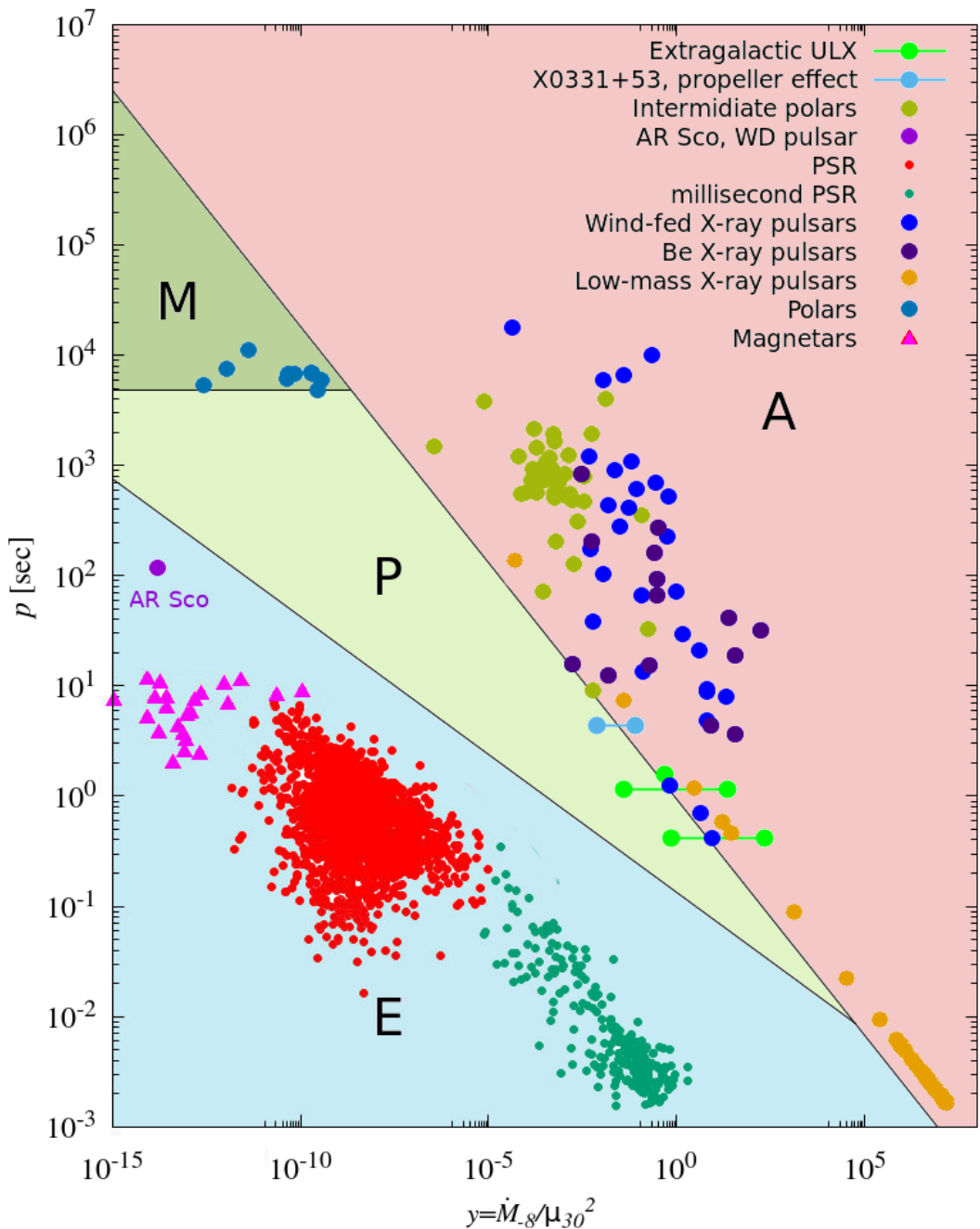

**Fig. 2.** Universal period-gravimagnetic parameters diagram for most of the observed types of neutron stars and white dwarfs. Intermediate Polars are shown in green; X-ray accreting pulsars, are ochre; ULX pulsating sources are in light green; radiopulsars with standard magnetic field ($\sim 10^{12}$ Gs) on neutron stars are in red, millisecond radiopulsars with low magnetic field ($\sim 10^{8-10}$Gs) on neutron stars are in green and Magnetars ($>10^{14}$Gs) are in magenta. Accreting X-ray Binary Pulsars: http://www.iasfbo.inaf.it/~mauro/pulsar_list.html . PSR: Manchester et al.,(2005); AR Sco: Marsh et al. (2016); X0331 + 53, Propeller: Raguzova & Lipunov (1998); Intermidiate Polars: Suleimanov et al. (2019); Polars: Cherepashchuk et al. (1996); Magnetars: Olausen & Kaspi (2014).

Upon careful consideration of the p-y diagrams, we notice that three groups of rotation-powered pulsars — millisecond, ordinary, and anomalous X-ray Pulsars (AXPs) — are distinguished not only by their position, but by the envelopes of the corresponding sets of straight lines (Fig. 3). We argue that this diagram surprisingly clearly shows the genetic relationship of X-ray pulsars in low-mass binary systems (X-ray sources of the Galaxy bulge) with millisecond pulsars. This connection, first proposed by several teams Alpar et al. (1982), Radhakrishnan and Srinivasan (1982) and by Fabian et al. (1983), is now generally accepted, but only the *p-y* diagram demonstrates it with all the evidence. Note that the idea of recycled pulsars actually came from Bisnovaty-Kogan and Komberg (1976). With an arrow we have shown how to stop the Roche Lobe red dwarf overflow , for example, a powerful radio pulsar evaporates with its relativistic wind its secondary component . Since the neutron star has almost lost its magnetic field, its spin down time is more than the Hubble time. As a result, millisecond pulsars mostly remember the magnetic field and the period established during accretion. This is the generally accepted scenario today for the formation of millisecond pulsars.

When switching to isolated standard radio pulsars (red dots), it is striking that the upper boundary of this subset of radio pulsars exactly repeats the slope of the line corresponding to the transition of the magnetic rotator from the ejector (**E**) to the propeller (**P**) state. On the one hand, this seems inevitable since the phenomenon of a radio pulsar is possible only in the vacuum region of the magnetosphere inside the light cylinder. On the other hand, it was noted by Kolosov et al. (1989) that the transition slope **E** → **P** coincides with the model of the line of death of radio pulsars in models of the formation of an electron-positron avalanche near the magnetic poles of neutron stars.

## 4. Equlibrium.

X-ray pulsars and intermediate polars, as expected, fall into the region of the Accretors. Obviously, to determine the y-parameter you need to know the accretion rate and the magnetic dipole moment. The accretion rate is determined from the observed X-ray luminosity $L_x = \dot{M}GM/R_x$. The radius and mass of the neutron star is everywhere assumed to be **10 km** and **1.5 $M_\odot$**.

For Intermediate Polars, we determine the magnetic dipole moment according to the observation of cyclotron lines and spectral features (Suleimanov et al., 2019); using the dipole formula $\mu = BR_x^3/2$. The radius for simplicity is assumed to be equal for all polars $R_x = 5000$ km .

And here, the magnetic field - we are for X-ray pulsars on neutron stars that have disk accretion realized from the condition that they are on average in equilibrium (Lipunov 1992):

$$\langle dI\omega/dt \rangle = \langle \dot{M}\sqrt{GMR_A} - \mu^2/R_c^3 \rangle = 0 \qquad (9)$$

Or

$$\langle \dot{M}\sqrt{GMR_A} \rangle = \mu^2 \omega_{eq}^2/GM \qquad (10)$$

From here we obtain the mean value of the gravimagnetic parameter *y*.

$$\langle y_{eq} \rangle \approx 10^{-42} \, p_{eq}^{6/7} \qquad (11)$$

Of the three principal parameters of the quantities included in these equations, $\dot{M}$, $\omega$ and $\mu$. the most strongly varying quantity is $\dot{M}$, so we can assume that $\omega_{eq} = 2\pi/p_{eq}$ that is, $p_{eq} \approx p_{obs}$, where $p_{obs}$ is the pulsar period observed at a given time.

This technique works well for Her X-1 type systems in which optical stars - donors have an average mass and accretion rate for which changes very little so that the pulsar most of the time is near the equilibrium value <$y_{eq}$>. This is proved by the fact that the mean acceleration time of a pulsar $t_{su}$ ≈ *300 000 years* is much longer than the strong acceleration observed in episodes $t_{su}$(min) ≈ 20 000 years (Lipunov 1987b) and therefore the pulsar is in the so-called "catastrophic equilibrium", see Fig. at page 147 of Lipunov (1992).

If we turn to dual systems with a red dwarf, this approach should already be used with caution. We are talking about donors with a mass of ~*0.3-0.7$M_\Theta$* with powerful convective layers below the surface. The flow of matter in such systems is extremely unstable and the rate of accretion may vary by orders of magnitude. Such a situation arises for rapidly evolving neutron stars - X-ray bursters or systems of neutron stars close to the formation of millisecond radio pulsars. And the same picture arises for intermediate polars. In these cases, the average value of <$y_{eq}$> ~ <$\dot{M}$> may differ by orders of magnitude from that observed at the moments of X-ray activity. In fact, the average accretion rate will be lower and the compact star most of the time can spend itself in the **Propeller (P)** or even the **Ejector (E)** phase. The *p-y* diagram (Fig. 2) for these sources shows the place corresponding to the episodes of X-ray activity.

A similar situation is realized for those X-ray accreting pulsars, where the donor is a rapidly rotating non-stationary massive star on the main sequence of the Be type. Here, the giant oscillations of the accretion rate are associated with the nonstationary character of the mass loss of a Be star due to the rotational instability of its outer layers. Due to the rapid rotation, such stars at the equator are surrounded by ejected disc-shaped shells. For dozens of years, Be stars can be in a state of calm, interrupted by the release of a shell. To this is added the significant ellipticity of the orbits of their neutron stars, which is another factor in the sharp variability of the accretion rate. The picture becomes truly chaotic and for these X-ray pulsars we show the *y*-parameter at the time of X-ray activity. Although in reality some of them, on average, are in a mixed state of **A↔P** Accretor-Propeller (Gnusareva et al.,1985).

**5. Neutron Stars Radiopulsars.**

Neutron stars were discovered as radio pulsars by Jocelyn Bell and Hewish (Hewish et al. 1968). As it should be, neutron-star radio pulsars all hit the Ejectors area **(E)**. Most of them are single neutron stars or stars paired with degenerate companions - so the donor here is the interstellar medium for which the average parameters for the Galaxy are accepted. The magnetic field of radio pulsars is found by measuring the slowing down rate, making the assumption that the energy losses of radio pulsars are described by the magnetic dipole formula: *$\mu_{30}$ = 10 (p $\dot{p}_{-15}$)$^{1/2}$*, where the period is measured in seconds, and the derivative of the period is dimensionless *$\dot{p}_{-15}$ = $\dot{p}/10^{-15}$* .

All radio pulsars are divided into two types - one (young neutron stars) are born immediately with small periods, also in binary systems. And others - old neutron stars spin up in the process of accretion in binary systems (recycled scenario, Alpar et al. (1982), Radhakrishnan and Srinivasan (1982) and Fabian et al. (1983)). On our diagram, these pulsars represent two areas: the first - at the left at the top of the Ejectors zone, and the second - the millisecond pulsars primarily belong to the millisecond pulsars - located in the lower right because of the small value of their magnetic field. Here we pay attention to the upper envelope of radio pulsars - first, the envelope goes parallel to the line (**P-E**) separating the Ejectors and Propellers. Physically, this is clear - the death stance of radio pulsars and the parallel line *p ~ $y^{-1/4}$* and possibly itself is the death line of not only ejectors, but radio pulsars (Kolosov et al.,1989).

Moving further along the upper edge, we notice that the slope increases and becomes parallel to the line separating the Propellers (P) and Accretors (A). This is natural since millisecond pulsars appear from old accreting stars lying on the equilibrium line, see Lipunov et al.,(1987a) (it was later called the Spin-Up Line).

**6. Magnetar, Soft Gamma Repeaters, Anomalous X-ray Pulsars and Fast Radio Bursts.**

The possibility of the existence of neutron stars with abnormally large magnetic fields was first suggested by (Woltjer, 1964) from simple magnetic flux conservation during evolution stars to dense state.

On March 5, 1979, the Soviet space gamma device Konus detected a bright gamma-ray burst (Mazets 1979) , which, along with a short bright flash, showed swell-out strictly periodic oscillations with a period of ~ 8 seconds. Hard spectrum with periodicity did not leave doubt, in that we are dealing with magnetized neutron star. The first time that the spin-down rate of Soft Gamma Ray Repeater was measured was in 1998 by Kouveliotou et al. (1998) for the SGR1806-20, which yielded a magnetic field strength approaching $10^{15}$ *G*. Later it turned out that similar neutron stars (anomalous X-ray pulsars : Ejectors) also show similar periods, and their change of periods corresponds to magnetic fields of $10^{14}$-$10^{15}$ *Gs*.

Now we know about 20 anomalous X-ray Pulsars (AXPs) and soft gamma repeaters (SGR) presented on our *p-y* diagram (Olausen, Kaspi 2014). All of them are in the left upper area of the Ejectors.

Last year the nature of FRB repeaters has finally become clear. The very phenomenon of millisecond flares in the radio range was predicted by Lipunov & Panchenko (1996). However, this prediction was about millisecond radio flares in the last millisecond before the merger of neutron stars. In fact, such unique impulses were discovered (Lorimer et al.,2007), however, along with them, repeating FRBs were also discovered. These FRB repeaters have been discussed for several years as manifestations of Magnetars, see Popov & Postnov (2013); Lyubarsky (2014); Beloborodov 2017; Metzger et al.,(2019).

In April of last year, for the first time, it was possible to register a radio burst from the galactic Magnetar of the Soft Gamma ray Repeater SGR / FRB 1935 + 2154 (Li et al.,2020; Mereghetti et al.,2020; Tavani et al.,2020; Ridnaia, A. et al.,2020 ; Bochenek et al.,2020).

Despite the large number of observations and theoretical works, the question of the main evolutionary factor of Magnetars still remains open. For anomalous X-ray pulsars, the rate of rotational energy losses corresponds to the magnetic dipole losses by magnetic fields of neutron stars ~ $10^{14-15}$ *Gs* (Kouveliotou et al. 1998), and they rightfully take their place among the ejectors in the diagrams Fig. 2. But even the most powerful Soft Gamma flares have no effect on the slowing down rate of neutron stars (Beloborodov 2020; Lyubarsky 2020). It seems that SGRs draw energy of the flares from the magnetic field (Beloborodov 2020). But in this case, the factors of the evolution of the magnetic rotator can be both a monotonic change in the rotation period (movement upward along the diagram Fig. 2) and dissipation of the magnetic field (movement to the right along the evolutionary diagram Fig. 3). In the latter case, as soon as the field drops to standard values, the neutron the star will cross the line of propellers and leave the Ejectors forever. The question of what is the main evolutionary factor for Magnetars as gravimagnetic rotators is still open.

There is no doubt only about one thing - Magnetars are ejectors with a characteristic lifetime of $10^{3-4}$ years. Below we will return to them and consider the possible evolutionary scenarios for their appearance.

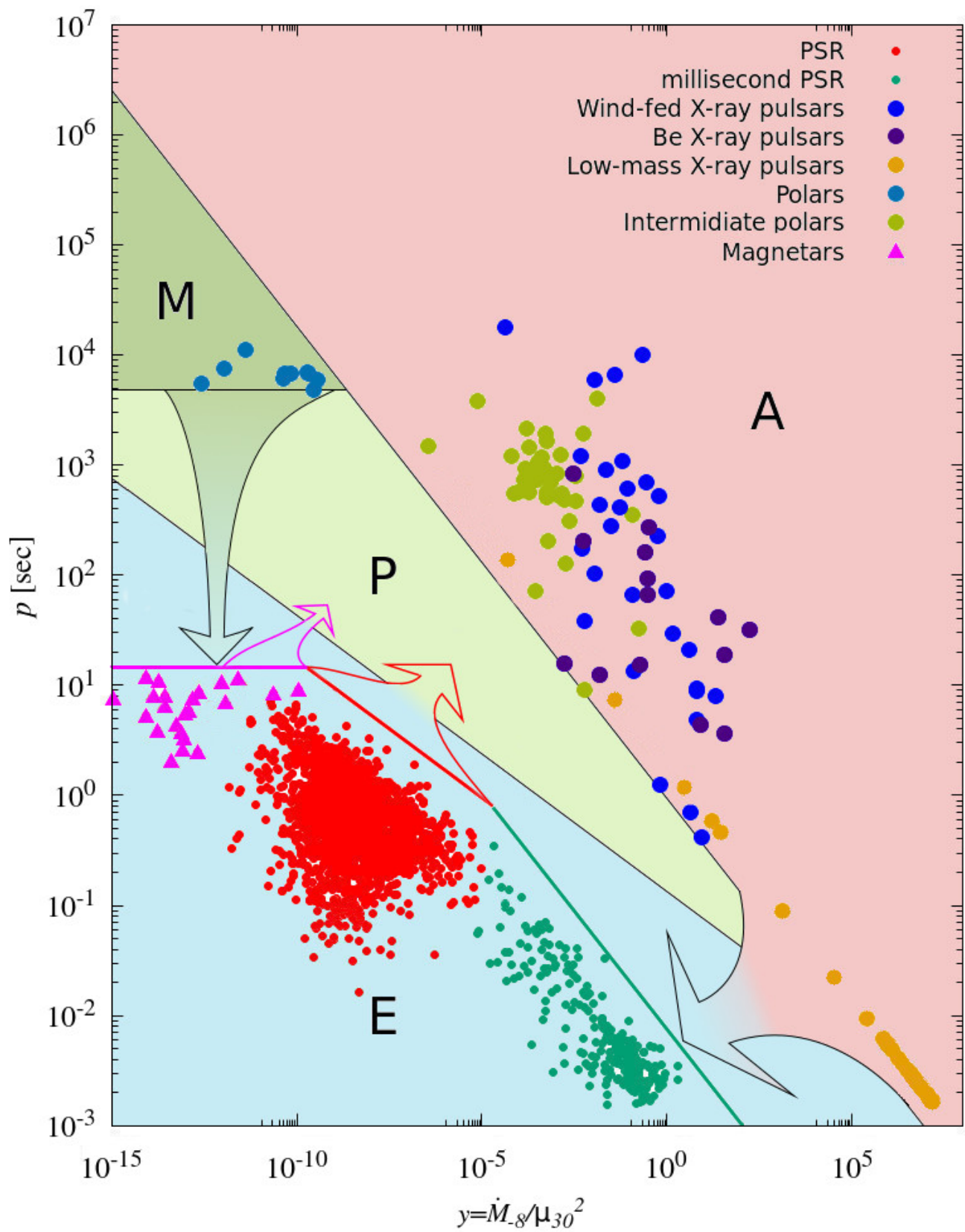

**Fig. 3.** Period-gravimagnetic parameters (*p-y*) diagram showing the suggested genetic relationship between Ejecting neutron stars (Magnetars, Radio Pulsars, Millisecond Pulsars) with accreting neutron stars and Polars.

## 7. Propellers

The source X0331 + 53b which shows the episodes of the propeller mode is depicted in Figure 2 by two blue dots corresponding to the transitions of the neutron star from the accretor state **(A)** to the propeller mode **(P)** and back. It should be emphasized that this transition is accompanied by a catastrophic (approximately 100 times) decrease in luminosity with a gradual weak change in the potential accretion rate, for example, due to the distance from the donor star in an eccentric orbit or consequence. As noted by Gnusareva et al. (1985) for the first time, this change is due to the passage through the centrifugal barrier. From the state $R_A < R_c$, the rotator enters the state $R_A > R_c$ and the accretion to the surface changes, and the total luminosity of the accretor $L_A = \dot{M}_c GM/R_0$ sharply (catastrophically) decreases to the luminosity of the Propeller $L_P = \dot{M}_c GM/R_A$. Since at the moment of the transition the accretion rate at the boundary of the magnetosphere changes smoothly, the magnitude of the jump (the amplitude of the X-ray dip (gap)) is: $L_P/L_A = R_x/R_A$. But at the moment of transition, the Alfven radius is approximately equal to the corotation radius, which is a function of the spin period of the GMR $R_c = (GM/\omega^2)^{1/3}$. So

$$L_P / L_A = R_x / R_c \approx 10^{-2} p^{-2/3} \qquad (12)$$

here we assumed $R_x = 10$ **km**, and the period is in seconds.

## 8. Ultra-Luminous X-ray Pulsars.

New in this diagram are pulsating ULX sources. We emphasize that these sources are not super-accretors, since the luminosity of the accretion flow in the magnetosphere does not exceed the Eddington limit and the entire plasma successfully reaches the surface of a neutron star, in the form of an accretion column above the magnetic poles. The X-rays are radiated in directions sideways from this column. As a result, the luminosity of such a X-ray pulsar can significantly increase the Eddington limit and reach the observed values for extragalactic objects whose pulsations have been discovered in recent years by **NGC 5907, NGC 7793 P13, M82 X-2** (see Table 2).

Table 2. Extragalactic ultraluminous X-ray sources

| Galaxy | P (s) | L (erg/s) | Satellite | Year and Reference |
|---|---|---|---|---|
| NGC 5907 | 1.42 | $2.2 \times 10^{41}$ | Swift, XMM, NuSTAR, Chandra | 2017 (Israel 2017a) |
| NGC 7793 P13 | 0.42 | $1.6 \times 10^{40}$ | Einsten, XMM, NuSTAR, Rosat | 2017 (Israel 2017b) |
| M82 X-2 | 1.37 | $3.7 \times 10^{40}$ | NuSTAR, Chandra | 2014 (Bachetti 2014) |

It is difficult to imagine another scenario in which the luminosity during stationary accretion (Kardashev 1964) on the magnetosphere would significantly exceed the local Eddington limit at $R=R_A$ and simultaneously the source would be an X-ray pulsar. This gives a natural limit to the accretion of subcritical regimes **A:**

$$y_{-42} \leq 10^7/\dot{M}_{18} \tag{13}$$

Recall that $y_{-42} = \dot{M}_{18}/\mu_{30}^2$. Putting $L_x = 10^{38} \dot{M}_{18}$ erg/s, we obtain that the condition of the subcritical mode of the accretor **A** can be written in the form:

$$L_x \leq 3 \cdot 10^{41} \mu_{30} \text{ erg/s} \tag{14}$$

At the magnetic poles of a neutron star, the luminosity in sideward directions can greatly exceed the Eddington limit. The radiation there can take on a sharply anisotropic character and mainly go across the accretion disk without touching the matter flows in the binary system, that is, without affecting the supercritical regime. In this case, the total luminosity of the pulsar can reach the value:

$$L_x = L_{Edd}(R_A/R_x) \approx 10^{40} \text{erg/s } y_{-42}^{-2/7} \tag{15}$$

$y_{-42} = y/10^{-42}$ **(CGS).** The sources mentioned above are present in the universal diagram as Accretors (**A**), and not Super Accretors (**SA**).

This mode is accompanied by significant losses in neutrino radiation, since in the polar column the temperature can exceed 1 Mev. The process of electron-positron pair production with energy loss through electro-weak interaction and creation neutrino and antineutrino (Zeldovich et al.,1972):

$$\gamma \rightarrow e^+ + e^- \rightarrow \nu_e + \nu_e^* \tag{16}$$

Such a scenario seems to be realized by ULX sources — pulsars, where the accretion rate can reach $10^{-6} - 10^{-5} M_\odot/yr$. This in turn gives rise to two new effects characteristic of these stages, which may be observed in the future (Lipunov 1992). First, since the luminosity of the polar column significantly exceeds the Eddington limit (Basco et al.,1976). Therefore, the accretion in the polar column must be pulsating, transient in nature at times (Lipunov 1992):

$$t \sim R_A/\sqrt{(GM_x/R_A)} \sim 2 \cdot 10^{-2} y_{-42}^{3/2} \text{ s} \tag{17}$$

Well, in a very distant future, we can expect the discovery of a neutrino pulsar caused by the anisotropy of neutrino production for neutron stars with an anomalously strong magnetic field $B > 4 \times 10^{13}$ Gs.

Returning to the X-ray sources in low-mass binaries, we note that they are located in the lower right part of the Accretor region. These are old neutron stars that have practically lost their magnetic field and sped up to the shortest periods of rotating objects in the Universe.

As for the ejectors, they are represented in the diagram by standard radio pulsars. The periods and dipole moments are taken from the catalog. Radio pulsars, and the accretion rate for all of them is taken as the average $M = 10^{10}$ g/s. The same applies to millisecond pulsars (the lower right corner of the **E**-zone), together with dual radio pulsars with degenerate component, as well as for Magnetars. This is understandable, since the interstellar medium is the donor for all these objects. An exception is the radio pulsar in a binary system with a blue (Johnston et al.,1992) Be star PSRB 1259-63.

To restore the accretion rate in this case is very difficult. But it is qualitatively clear that each orbital revolution in this system leads to a periodic change in the *y* parameter, which we demonstrate with a horizontal segment.

Of particular interest is the newcomer among the ejectors of the radio pulsar - the white dwarf AR Sco!

## 9. Polars.

Finally, we will look at the new class of these objects in the p-y diagram. In the process of preparing this paper, it became clear to us that the Polars can be naturally shown in this diagram in a separate area - the field of Magnetors (M). This type of magnetosphere was proposed by Mitrofanov et al. (1977) as accreting white dwarfs with magnetic fields so strong that the donor( an optical star) is inside the white dwarf magnetosphere: $R_A > a$.

But the Alfven radius is determined by the *y* parameter, and the semi-axis is double, respectively:

$$R_A = (2y\sqrt{2GM_{wd}})^{-2/7}, \quad a = (G(M_o+M_{wd})/4\pi^2)^{1/3} \, p^{2/3} \tag{18}$$

Assuming the sum of the masses characteristic of binary systems - the polar consisting of red and the white dwarfs is equal to 1 $M_\odot$. Equating both values, we note that the critical line for Magnetors - that is, the polar.

$$p_M \geq 2\pi/(GM_{tot})^{-1/2} \, a^{3/2} \approx 10^3 \text{ s } (M_{tot}/0.1M_\odot)^{-1/2}(a/0.1R_\odot)^{3/2} \tag{19}$$

where *a* - is binary separation.

We drew a horizontal line for $M_{tot} = 0.5M_\odot$, $R = 0.1 R_\odot$. However, unlike the Accretor, the Magnetor period is determined by the magnetic synchronization of white and red dwarfs and is exactly equal to the orbital period of the binary system. In other words, Magnetors - Polars should line up along the $p_M$ line (Cherepashchuk et al.,1996).

## 10. Polars as the Magnetars progenitor?

Now we return to the third group of neutron stars, which are currently called anomalous X-ray pulsars (AXP) and Soft Gamma-ray Repeaters (SGRs). At present, all of them are interpreted as Magnetars — neutron stars with an anomalously strong magnetic field of ~ $10^{14} - 10^{15}$ Gs at the ejection stage (**E**). The nature and mechanisms of energy release are far from clear. In the *p-y* diagram, they are shown as pink triangles in the assumption that they are all at the Ejector stage (**E**).

After the discovery of the Magnetar CXOU J1664710.2-455216 in the young star cluster Westerlund 1 (Clark et al. 2014; Ritchie et al. 2010), the point of view in the scientific community that all Magnetars (Anomalous X-ray Pulsars - AXP, Soft Gamma Repeaters - SGR, and Fast Radio Bursters) are descendants of very massive, short-lived stars. This scenario can be called phenomenological, that is, a scenario in which the hypothesis of the formation of highly magnetized neutron stars is simply a hypothesis about the initial conditions: there are massive stars producing neutron stars with an abnormally strong magnetic field. Of course, at the same time, words are spoken about the fact that the cores of these stars, before the collapse, already have a sufficiently large magnetic flux and fast rotation. In this case, the origin of fast rotation or large magnetic flux is postulated. It is assumed that due to the dynamo mechanism, the magnetic field strength on the surface of the forming stars reaches $10^{15}$ Gs, and in some recent works, fields of $10^{16-18}$ *Gs* are considered (Beloborodov 2020). However, in reality, there are at least three more situations when the formation of neutron stars is accompanied by an initial super-powerful magnetic flux or rapid rotation, which leads to an increase in the initial magnetic field.

In one such scenario, a Magnetar is formed by merging neutron stars (Giacomazzo & Perna, 2013; Piro & Kollmeier, 2016). In this case, there is a clear genetic relationship with short GRBs (Abbott et al., 2017; Lipunov & Gorbovskoy, 2008; Lipunov et al.,2018) and possibly with Fast Radio Bursts (Lipunov & Panchenko, 1996). Here, a large angular momentum arises from the orbital motion in the binary. An even greater angular momentum can be realized in the case of merging of heavy oxygen-neon-magnesium (O-Ne-Mg) white dwarfs with masses close to the Chandrasekhar limit (Lipunov, 2017).
This process occurs at approximately the same frequency as the merger of neutron stars and can be accompanied by a powerful gravitational-wave impulse of a non-spiral wave form (Lipunov, 2017). Recall, accretion induced collapse (AIC) of O-Ne-Mg white dwarfs was already predicted by Miyaji et al. (1979).

Finally, there is a scenario for the formation of highly magnetized neutron stars as a result of accretion induced collapse (Lipunov & Postnov, 1985; Ruiter et al.,2019). Let us consider in more detail the scenario of the formation of Magnetars from other, anomalously magnetized objects - white dwarfs with magnetic fields of more than 100 MGs. Such white dwarfs have long been observed in close binaries and are called Polars. The most massive of them can gain mass and, after reaching the Chandrasekhar limit, transform into a neutron star with a high magnetic field. This scenario was proposed by Lipunov & Postnov (1985) for the first anomalous X-ray pulsar 1E2259 + 59. The last two scenarios have received unexpected support recently. In 2020, Magnetar was discovered in a galaxy with weak star formation, in which all massive objects have long died (Ruiter et al.,2019).

We notice, that the envelope of the millisecond, ordinary and anomalous X-ray pulsar line here is sharply bent and becomes almost horizontal. At the same time, the Polars behave in a similar manner with respect to their period, as a result of magnetic synchronization, which is simply determined by the third Kepler law. It seems tempting to draw on the old hypothesis of the formation of anomalous X-ray pulsars as descendants of the Polar like binary systems. We propose that during accretion process abnormally magnetized white dwarfs gain mass and collapse to neutron stars after reaching the Chandrasekhar limit. In the approximation of conservation of angular momentum $I\omega = const$, and magnetic flux, the period of the formed neutron star and its dipole magnetic moment can be estimated as:

$$p_{NS} = p_{WD} (I_{NS}/I_{WD}) \approx 10^{-3} \ p_{WD} \approx 10s \tag{20}$$

$$\mu_{NS} = \mu_{WD} (R_{NS}/R_{WD}) \approx 10^{35} (10 \text{ km}/700\text{km}) \ Gs \ sm^3 \approx 1,5 \ 10^{34} \ Gs \ cm^3 \tag{21}$$

since the white dwarf inertia moment is small near the Chandrasekhar limit. In this scenario, anomalous X-ray Pulsars (AXPs) and gamma repeaters periods are explained naturally by their abnormally large magnetic moments.

Next, we modeled millions Polars and using the conservation laws, we obtained millions neutron star - Magnetars from them. The resulting parameters for the neutron stars are as follows.

About seven hundred Polars are known, but not for all of them all the parameters are known. Fig. 4. and Fig. 5 shows a histogram of known Polar periods and magnetic field (http://physics.open.ac.uk/RKcat/cbcat)

However, these are just the initial parameters of the Magnetars being born. We assume that all anomalous phenomena observed in Magnetars just occur at the beginning of their birth. This is confirmed by the very small age of $\sim 10^{3-5}$ yrs attributed to these objects (Mereghetti et al.,2015). In turn, their age is independently confirmed by the presence of supernova remnants near these objects, whose age is of the same order!

We emphasize that this scenario does not mean that all Polars turn into Magnetars. The fact is that usually the collapse of a carbon white dwarf is accompanied by a complete explosion of the white dwarf (type Ia supernova). However, hydrodynamic calculations demonstrate that heavy white dwarfs, so-called oxygen-neon-magnesium ones, collapse into a neutron star (see Miyaji et al. (1979), Nomoto et al. (1984), Tominaga et al. (2013)) . These heavy white dwarfs are born as a result of the evolution of sufficiently massive progenitors with a mass in the range of **$M_0 \sim$ 8-10 $M_\odot$**.

Such stars form in accordance with the Salpeter initial mass function:

*dN/dM ~ M^-α*                                                                                                                                    (22)

Where (Fedorova et al.,2004) **α ≈ 2.35** .

On the other hand, SNe Ia are associated with heavy CO-white dwarfs, which arise from progenitors of the order of more than 3 solar mass, that is, about $\sim$ 10 times more often. The frequency of Ia supernova explosions is $\sim$ 1/300 yrs$^{-1}$ in our Milky Way galaxy. So we can state that the birth rate of Magnetars in this scenario is of the order of $\sim$ 1/1000 - 1/3000 yrs$^{-1}$. Multiplying by average age of anomalous X-ray Pulsars (AXPs) ($10^{4-5}$ yrs), we get their total number in the galaxy $\sim$ 10-30, which is in accordance with the observed satitistics of them (Mereghetti et al.,2015, Bisnovaty-Kogan et al.,2014, Olausen et al.,2014) .

It is easy to make the bridge to the Fast Radio Burst (**FRB**) in this scenario. As it's known, FRBs are short (at several tens of milliseconds) radio flashes occurring at cosmological distances. Nowadays these objects are regularly recorded by the radio telescopes and we have sufficiently accurate statistics of these phenomena which agrees with events in a galaxy of our $10^{-3}$ yrs$^{-1}$ type. Short radio bursts were predicted (Lipunov & Panchenko 1996) as original precursors of neutron star merging. However, repeating FRBs do not agree with the hypothesis of neutron star merger. Also, realistic estimates of the NS merging frequency even at cosmological distances are much smaller than the observed frequency of FRB events. This likely means that the FRB events are a mixture of different physical processes, among which prevailing in frequency is not a merger but super-flare on Magnetars and gamma-ray repeaters (Katz 2018) .

## 11. Population synthesis of neutron stars' formation from intermediate mass binary systems.

In this section, we will simulate the formation of neutron stars from heavy O-Ne-Mg Polars using the laws of conservation of torque and magnetic flux. Then we check the parameters of the generated neutron stars and filter out from them that are not Magnetars.

Now about seven hundred Polars are known, but not for all of them all of their parameters known. Fig. 4 shows a histogram of the known periods and values of the magnetic field strength of Polars (Ferrario, 2015), as well as a histogram of the modeled Polars. We used the distribution of visible Polars and approximated it with a lognormal distribution.

We will assume that the laws of conservation of angular moment and magnetic flux are fulfilled during the collapse of Polar (WD) into a Neutron star (NS):

$$P_{NS} = P_{WD} \cdot I_{NS}/I_{WD} \tag{23}$$

$$\frac{\mu_{NS}}{R_{NS}} = \frac{\mu_{WD}^{ONeMg}}{R_{WD}^{ONeMg}} \tag{24}$$

Where $\mu = BR^3/2$ – is the magnetic dipole moment.

Since O-Ne-Mg white dwarfs are quite rare, we will assume that all the polars we see are C-O white dwarfs. Thus, assuming that the magnetic flux of both types of white dwarfs is the same, we rewrite (24) as follows:

$$\frac{\mu_{WD}^{ONeMg}}{R_{WD}^{ONeMg}} = \frac{\mu_{WD}^{CO}}{R_{WD}^{CO}} \tag{25}$$

We will take the following parameters in the calculations:

$R_{WD}^{ONeMg} = 7 \cdot 10^7$ cm. – is O-Ne-Mg WD radius (WD)
$R_{WD}^{CO} = 7 \cdot 10^8$ cm. – is C-O WD radius.
$R_{NS} = 10^6$ cm. – Magnetar radius (NS)
$k_{WD} = 0.0754$ – is the radius of gyration
$M_{WD} = 1.4 \cdot 2 \cdot 10^{33}$ g. – is WD mass at chandrasekhar limit
$R_{NS}^G = 3.6 \cdot 10^5$ cm. – is gravitational radius of NS
$M_{NS} = M_{WD} (1-(R_{NS}^G/(2 \cdot R_{NS})))$ g. – is Magnetar mass

We calculate the moments of inertia for Polar (WD) and neutron star (NS) using the well-known formula:

$$I_{WD} = k_{WD} M_{WD} R_{WD}^2 \tag{26}$$
$$I_{NS} = 2/5 \cdot M_{NS} \cdot R_{NS}^2 \tag{27}$$

The law of deceleration of the periods of Magnetars:

$$\frac{dI\omega}{dt} = -\chi \left(\frac{\mu^2 \cdot \omega^3}{c^3}\right) \tag{28}$$

Restrictions accepted for Magnetars:

$B_{Crit} = 0.45 \cdot 10^{14}$ G. – critical magnetic field

$\tau_{max} = 10^4 - 10^6$ yr. – Magnetar lifetime, $\chi = \frac{2}{3}$.

We also take into account that the characteristic age of the Magnetar $\tau = P/2\dot{P}$. We assume that all neutron stars with $B_{NS} < B_{Crit}$ or $\tau < \tau_{max}$ have not ceased to be Magnetars. Thus, we calculate the distributions of the obtained distributions of Magnetars over magnetic fields and periods. We will also take into account that the Magnetars we observe live and are born in their galaxy are in a steady state such that their number is independent of time. During $\sim (3 - 4) \cdot 10^5$ years, a steady-state population of Magnetars is then established.

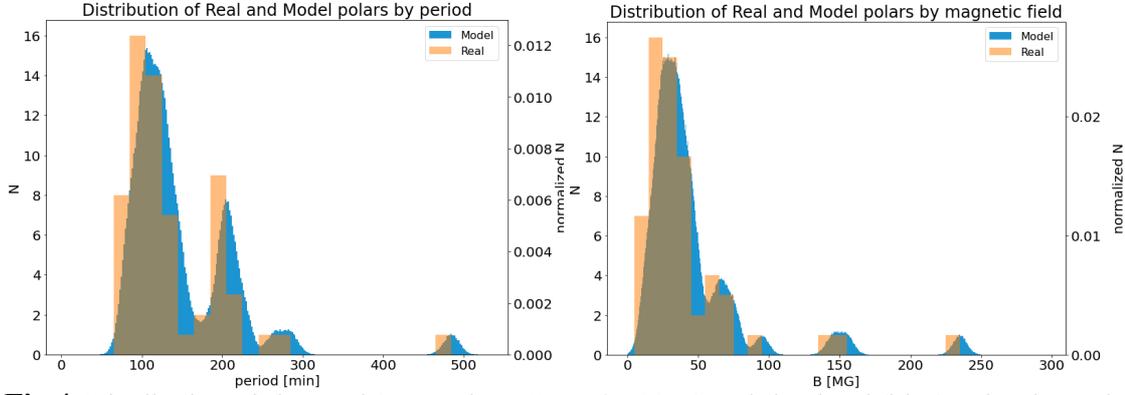

**Fig.4.** Distribution of observed (orange bars, Ferrario, 2015) and simulated (blue) polars by period (left panel) and by magnetic field strength (right panel). The areas of both distributions are normalized to 1.

Fig. 5 shows the calculation results for the initial distribution of polars in comparison with visible Magnetars (orange histogram, McGill Online Magnetar Catalog[1]). Fig. 5 shows the calculation for the initial distribution of polars in comparison with the distribution of observed Magnetars (orange histogram, field. However, we cannot exclude that white dwarfs with even larger magnetic fluxes may be discovered in the future.

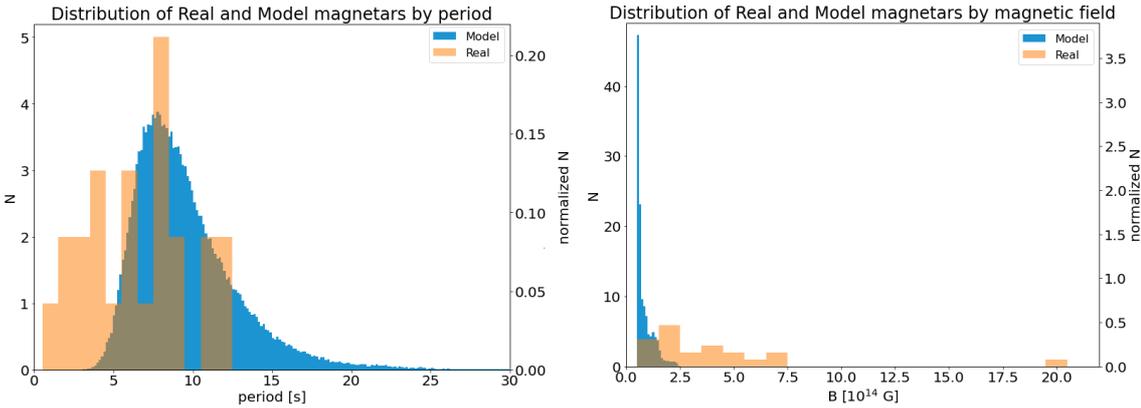

**Fig. 5.** Distribution of observed Magnetars (orange bars) and model Magnetars obtained from polars (blue) by periods (left panel) and by magnetic field strength (right panel). The areas of both distributions are normalized to 1.

---

[1] http://www.physics.mcgill.ca/~pulsar/Magnetar/main.html

## 12. Conclusion.

We mainly focused on the most studied magnetized astrophysical objects for which the Gravity Magnetic Rotator model can be applied. These are primarily neutron stars and white dwarfs.

Despite its approximate form, the "p-y" diagram not only confirmed the generally accepted genealogy of millisecond pulsars–the oldest neutron stars (aged ~$10^{10}$yrs), but also the fading mechanisms of standard intermediate-age pulsars (~$10^{6-7}$yrs). We have discussed 4 possible branches of Magnetars formation. For one of the scenarios, we performed a small population synthesis calculation, which showed good agreement with the observed magnetic fields and Magnetar periods. In this scenario, Magnetars can be born in elliptical galaxies and galaxies with weak star formation. This becomes especially relevant after the discovery of Magnetars in galaxies with weak or no star formation.

When our work was completed, an article by Kirsten et al.2021, was published, which reports on the localization of FRB 20200120 in one of the globular clusters of the galaxy M81. And after the discovery of a FRB originating from soft gamma-ray repeater SGR 1935 + 2154 (Li et al.,2020), this shows that the accretion-induced collapse scenario of magnetic white dwarfs (Lipunov & Postnov, 1985), considered in detail in this work, is an actual genealogical branch of Magnetar production.

The future development of the GMR model can be seen from two different points of view. The first one is the more adequate and detailed description of the individual modes of interaction of GMR with the surrounding plasma. In fact, that will give a good result only for some classes of objects, for example, if we examine in detail a particular stage in the evolution of a binary system. For example, cataclysmic variables, their mass distributions, orbital periods, etc. However, it should be borne in mind here that various uncertainties of the evolutionary scenarios (such as an adequate description of the common envelope stage, the Roche lobe overflow process, magnetic stellar wind parameters) could hardly have made the detailed calculations substantially more accurate than they are already (i.e. they remain undefined with the same coefficient of order 2). On the other hand, we are counting on progress in the study of extragalactic binary systems.

**Acknowledgements**
Authors are grateful to E.P.J. van den Heuvel, A.Tutukov, N.Tyurina, G.Lipunova, E.Gorbovskoy and K.Zhirkov for the fruitful discussion. Also authors would like to thank the referee for the useful comments, that helped to improve the article.
The presented work is supported in parts by Lomonosov Moscow State University Development program and by RFBR grant 19-29-11011.

**REFERENCE**

Abbott, A., et al.,2017. Multi-messenger Observations of a Binary Neutron Star Merger. ApJ, 848, L12 . 10.3847/2041-8213/aa91c9
Alpar, A.., Cheng, A..F., Ruderman, M.A., Shaham, J., 1982. A new class of radio pulsars. Nature, 300, 728-730 https://doi.org/10.1038/300728a0
Amnuel, P.R., Guseinov, O., H., 1968. Xray emission during interstellar medium accretion by neutron star. Notices (Izvestia) of Azerbadzhan Academy of Science, Physics and Mathematic series, 3, 70-74

Basko, M.M. , Sunyaev, R.A., 1975. Radiative transfer in a strong magnetic field and accreting X-ray pulsars. A&A, 42, 311 -321 http://adsabs.harvard.edu/abs/1975A%26A....42..311B

Beloborodov, A. M., 2017. A Flaring Magnetar in FRB 121102? ApJL, 843, L26 https://doi.org/10.3847/2041-8213/aa78f3

Bisnovatiy-Kogan, G.S., Friedman A.M., 1969. A Mechanism for Emission of X Rays by a Neutron Star. AZh, 46, 721  https://ui.adsabs.harvard.edu/abs/1969AZh....46..721B/abstract

Bisnovaty-Kogan, G.S., Komberg, B.V., 1976. Possible evolution of a binary-system radio pulsar as an old object with a weak magnetic field. SvAL, 2, 130-132 https://ui.adsabs.harvard.edu/abs/1976SvAL....2..130B/abstract

Beloborodov, A., 2020. The American Astronomical Society, find out more  The American Astronomical Society, find out more  The Institute of Physics, find out more  The Institute of Physics, find out more  Blast Waves from Magnetar Flares and Fast Radio Bursts. ApJ, 896, (2), id.142  https://doi.org/10.3847/1538-4357/ab83eb

Bochenek, C. D., Ravi, V., Belov, K. V. et al 2020. A fast radio burst associated with a Galactic Magnetar, Nature, 587, 59-62
 https://doi.org/10.1038/s41586-0

Beloborodov, A., 2020. Blast Waves from Magnetar Flares & Fast Radio Bursts. ApJ, 896, 2, 142.
 10.3847/1538-4357/ab83eb

Bondi, H., 1952. On Spherically Symmetrical Accretion, MNRAS,112, 195-204. 10.1093/mnras/112.2.195

Bondi, H. and Hoyle F.,1944; On the mechanism of accretion by stars,  MNRAS, **104**, 273-282.

Cherepashchuck et al., 1996 , Highly evolved close binary stars, Amsterdam : Gordon and Breach, QB821, H53 1996, 1, 386. http://adsabs.harvard.edu/full/1995SSRv...74..313C

Clark, J. S. et al.,2014. A VLT/FLAMES Survey for Massive Binaries in Westerlund 1. A&A, 565, A90. Ihttps://doi.org/10.1051/0004-6361/201321771

Crawford, J.A., 1955. On the subgiant components of eclipsing binary systems ApJ, 121, 71. https://doi.org/10.1086/145965

Davidson K. and Ostriker J.P., 1973, Neutron star accretion in a stellar wind: model for a pulsed X-ray sources, Ap.J., 179,585-598. 10.1086/151897

Davies, R. E.; Pringle, J. E., 1981. Spindown of neutron stars in close binary systems. MNRAS, 196, 209D https://doi.org/10.1093/mnras/196.2.209

Davies, R.E., Pringle J.E., 1980. On accretion from an inhomogeneous medium. MNRAS, 191, 599 https://doi-org.eres.qnl.qa/10.1093/mnras/191.3.599

Johnston ,S., Manchester , R.N., Lyne, A.G. et al.,1992.  PSR 1259-63: A Binary Radio Pulsar with a Be Star Companion . ApJ, 387, L37 10.1086/186300

Fabian, A.C. , Pringle, J.E., Verbunt, F., Wade, R.A., 1983.  Do galactic bulge X-ray sources evolve into millisecond pulsars? Nature, 301,222-223. 10.1038/301222a0

Ferrario, L., de Martino, D. & Gänsicke, B.T., 2015 Magnetic White Dwarfs . Space Sci Rev, 191,  111–169 10.1007/s11214-015-0152-0

Hakan Erkut M., Ekşi K.Y., Alpar. M.A., 2019 Ultra-luminous X-Ray Sources as Super-critical Propellers ApJ,873,105E 10.3847/1538-4357/ab04ae

van den Heuvel, E.P.J. , Heise, J. , 1972. Centaurus X-3, possible reactivation of an old neutron star by mass exchange in a close binary.Nature. Phys. Sci. 239, 67 https://doi.org/10.1038/physci239067a0

van den Heuvel, E.P.J.,  1977. Evolutionary process in X-ray binaries and their progenitor systems Ann. New York Acad. Sci., 302, 13 10.1111/j.1749-6632.1977.tb37034.x

Hewish A., Bell  S. J., Pilkington  J. D. H., Scott  P. F., Collins, R. A. 1968. Observation of a Rapidly Pulsating Radio Source.  Natur, 217, 709H 10.1038/217709a0

Giacomazzo, B. & Perna, R., 2013. Formation of stable Magnetars from binary neutron star mergers. ApJ 771, L26. 10.1088/2041-8205/771/2/L26

Gnusareva,V.S., Lipunov, V.M.,1985. Neutron Star Evolution in Eccentric Orbit Binaries . Soviet Astronomy, 29, 645-649 http://adsabs.harvard.edu/abs/1985SvA....29..645G

Gold, T., 1968. Rotating Neutron Stars as the Origin of the Pulsating Radio Sources. Nature, 218, 731G 10.1038/218731a0

Giacconi, R., Murray, S.et al. 1972. The Uhuru catalog of X-ray sources. ApJ, 178, 281G 10.1086/151790

Illarionov A. F., R.Sunyaev, 1975. Why the Number of Galactic X-ray Stars Is so Small? A&A, v.39, 185 https://ui.adsabs.harvard.edu/abs/1975A%26A....39..185I/abstract

Israel, G.L., 2017. et al.,An accreting pulsar with extreme properties drives an ultraluminous x-ray source in NGC 5907. Sci,355,817I 10.1126/science.aai8635

Israel, G.L. et al., 2017. Discovery of a 0.42-s pulsar in the ultraluminous X-ray source NGC 7793 P13 MNRAS.466L..48I 10.1093/mnrasl/slw218

Johnston, S., Manchester R.N., Lyne A.G. et al., 1992. PSR 1259-63 - A binary radio pulsar with a Be star companion ApJ, 387, L37 10.1086/186300

Kardashev N.S., 1964. Magnetic Collapse and the Nature of Intense Sources of Cosmic Radio-Frequency Emission AZh, 41, 807K ( translation in 1965, SvA, 8, 643K )
https://ui.adsabs.harvard.edu/abs/1964AZh....41..807K/abstract
https://ui.adsabs.harvard.edu/abs/1965SvA.....8..643K/abstract

Katz, J.L., 2018. Are Fast Radio Bursts Made By Neutron Stars? MNRAS Letters, 494, 1, p.L64-L68 https://ui.adsabs.harvard.edu/link_gateway/2020MNRAS.494L..64K/doi:10.1093/mnrasl/slaa038

Kirsten, F. et al.,2021. A repeating fast radio burst source in a globular cluster. arXiv:2105.11445v1 https://arxiv.org/abs/2105.11445

Kolosov, D. E., Lipunov, V. M., Postnov, K. A., Prokhorov, M. E.,1989. Recycled radiopulsar reservation in the P-P(dot) diagram. A&A, 215, 2, p.L21-L23.
https://ui.adsabs.harvard.edu/abs/1989A%26A...215L..21K/abstract

Kumar, P., Lu, W. & Bhattacharya, M. Fast radio burst source properties and curvature radiation model. 2017. MNRAS, 468, 2726–2739. https://10.1093/mnras/sty2829

Li, C. K. et al.,2020. Identification of a non-thermal X-ray burst with the Galactic Magnetar SGR 1935+2154 & a fast radio burst with Insight-HXMT. Preprint at https://arxiv.org/abs/2005.11071

Lipunov, V.M., 1982a. The universal diagram for magnetized neutron stars in the galaxy. Ap&SS, 85, 451L http://adsabs.harvard.edu/full/1982Ap%26SS..85..451L

Lipunov, V.M., 1982b. Supercritical Disk Accretion onto Magnetized Neutron Stars. Soviet Astronomy, v. 26, 54-56 http://adsabs.harvard.edu/abs/1982SvA....26...54L

Lipunov, V.M., 1982c. The magnetic fields of X-ray pulsars. Soviet Astronomy, 26, 537-541 https://ui.adsabs.harvard.edu/abs/1982SvA....26..537L/abstract

Lipunov, V.M., 1987a. The Ecology of Rotators . Astrophysics and Space Science, 132, 1-50, https://ui.adsabs.harvard.edu/abs/1987Ap%26SS.132....1L/abstract

Lipunov, V.M., 1987b. On the Equilibrium of HERCULES-X-1. Soviet Astronomy, V.31 (2), 167 https://ui.adsabs.harvard.edu/abs/1987SvA....31..167L/abstract

Lipunov,V.M., 1992a. Astrophysics of Neutron Stars, Springer-Verlag, Berlin, Heidelberg, New York. Also Astronomy and Astrophysics Library. https://www.springer.com/gp/book/9783642763526

Lipunov, V.M., 2017. Double O-Ne–Mg white dwarfs merging as the source of the powerfull gravitational waves for LIGO/VIRGO type interferometers. NewA, 56, 84L, https://doi.org/10.1016/j.newast.2017.04.013

Lipunov, V.M., Gorbovskoy, E., 2007. An Extra Long X-Ray Plateau in a Gamma-Ray Burst and the Spinar Paradigm. ApJ, 665, L97 10.1086/521099

Lipunov, V.M., Gorbovskoy, E.S., 2008. Spinar paradigm and the central engine of gamma-ray bursts. MNRAS, V. 383, 1397 https://doi:10.1111/j.1365-2966.2007.12489.x

Lipunov, V.M. Kornilov, V. G. Gorbovskoy, E. et al.,2018. The discovery of the neutron stars merger GW170817/GRB170817A and a binary stars evolution . NewAst, 63, 48-60. 10.1016/j.newast.2018.02.004

Lipunov, V.M., Nazin, S.N., 1992. Recycled white dwarfs?. Astronomical and Astrophysical Transactions, 1 (2), 153-157 10.1080/10556799208244528

Lipunov, V.M., Panchenko, E., 1996. Pulsars revived by gravitational waves. Astron. Astrophys, 312, 937-940. https://ui.adsabs.harvard.edu/abs/1996A%26A...312..937L/abstract

Lipunov, V.M., Postnov, K.A., 1985. The binary X-ray pulsar 1E 2259+59 - a descendant of an AM HER type system ? A&A, 144, L13-L14 . https://ui.adsabs.harvard.edu/abs/1985A%26A...144L..13L/abstract

Lipunov, V.M., K.A.Postnov, 1988. The Joint Evolution of Normal and Compact Magnetized Stars in Close Binaries - Analytical Description and Statistical Simulation. Astrophysics and Space Science, 145, 1-45 10.1007/BF00645692

Lipunov, V.M., Postnov, K.A., Prokhorov, M. E., Panchenko, I.E., Jorgensen, H.E.,1995. Evolution of the Double Neutron Star Merging Rate and the Cosmological Origin of Gamma-Ray Burst Sources . ApJ, 454, 593L. 10.1086/176512

Lipunov, V.M., Postnov, K.A., Prokhorov, M. E. 1996. The Scenario Machine: Binary star population synthesis. Amsterdam: Harwood Academic Publishers (http://xray.sai.msu.ru/~mystery/articles/review/ ) http://adsabs.harvard.edu/abs/1996smbs.book.....L

Lipunov, V. M., Postnov, K.A., Prokhorov, M. E., Bogomazov, A. I., 2009. ARep, 53 (10), 915-940 http://adsabs.harvard.edu/abs/2009ARep...53..915L

Lipunov V.M. et al.,2016. IceCube HESE 160814: MASTER OT J130845.02-323254.9 as the possible source of the high energy neutrino. GCN Circular, 19888, 1 https://gcn.gsfc.nasa.gov/gcn3/19888.gcn3

Lipunov, V.M., Shakura, N.I., 1976. SvAL, 2, 133L http://adsabs.harvard.edu/abs/1976SvAL....2..133L

Lipunova, G.V., 1997. A burst of electromagnetic radiation from a collapsing magnetized star. AstL, 23(1), 84-92 http://adsabs.harvard.edu/abs/1997AstL...23...84L

Lipunova, G.V., Lipunov V.M, 1998. Formation of a gravitationally bound object after binary neutron star merging and GRB phenomena . A&A, 329, L29-L32 https://ui.adsabs.harvard.edu/abs/1998A%26A...329L..29L/abstract

Lipunova, G.V., Gorbovskoy, E. S., Bogomazov, A.I., Lipunov, V.M., 2009. Population synthesis of gamma-ray bursts with precursor activity and the spinar paradigm . MNRAS, 397, 1695-1704 https://ui.adsabs.harvard.edu/abs/2009MNRAS.397.1695L/abstract

Lyubarsky, Yu., 2014. A model for fast extragalactic radio bursts. MNRAS. 442, L9–L13 10.1093/mnrasl/slu046

Lyubarsky, Yu., 2020. Fast radio bursts from reconnection in Magnetar magnetosphere. . Preprint at https://arXiv:2001.02007

Manchester, R. et al. 2005. The Australia Telescope National Facility Pulsar Catalogue. AJ, 129, 1993-2006 http://adsabs.harvard.edu/abs/2005AJ....129.1993M

Marsh T.R.et al., 2016. A radio-pulsing white dwarf binary star. Nature, 537, 374–377 https://doi.org/10.1038/nature18620

Mazets, E. P., Golenetskii, S. V. et al., 1979. Observations of a flaring X-ray pulsar in Dorado. Nature, 282, 587–589 https://doi.org/10.1038/282587a0

Mereghetti, S., Pons, J., Melatos, A., 2015. Magnetars: Properties, Origin and Evolution. Space Science Reviews, 191 (1-4), 315 10.1007/s11214-015-0146-y

Mereghetti, S. et al., 2020. INTEGRAL discovery of a burst with associated radio emission from the Magnetar SGR 1935+2154. ApJ Lett. 898, 29 https://doi.org/10.3847/2041-8213/aba2cf

Metzger, B. D., Margalit, B. & Sironi, L., 2019. Fast radio bursts as synchrotron maser emission from decelerating relativistic blast waves. Mon. Not. R. Astron. Soc. 485, (3), 4091–4106. https://doi.org/10.1093/mnras/stz700

Miyaji, S. ; Sugimoto, D. ; Nomoto, K. ; Yokoi, K.,1979. Supernova Explosion Triggered by Electron Capture and Formation of a Neutron Star ICRC, 2, 13M https://ui.adsabs.harvard.edu/abs/1979ICRC....2...13M/abstract

Mitrofanov, I.G., Pavlov, G.G., Gnedin, Yu.,1977. The nature of the AM Herculis system /3U 1809+50. SovAstL, 3, 182-184 https://ui.adsabs.harvard.edu/abs/1977PAZh....3..341M/abstract

Olausen S. A., Kaspi, V.M., 2014. The McGill Magnetar Catalog. ApJS,212, 6 10.1088/0067-0049/212/1/6

Nomoto,K., 1984. Evolution of 8-10 solar mass stars toward electron capture supernovae. I - Formation of electron-degenerate O + NE + MG cores. ApJ, 277, 791-805 10.1086/161749

Pacini F.. 1967 Energy Emission from a Neutron Star. Nature, 216, 567 https://doi.org/10.1038/216567a0

Piro, A. L. & Kollmeier, J. A. Ultrahigh-energy cosmic rays from the en caul birth of Magnetars. ApJ 826, 97 (2016) https://doi.org/10.3847/0004-637X/826/1/97

Popov, S. B., Postnov, K. A., 2013. Millisecond extragalactic radio bursts as Magnetar flares. Preprint at https://arXiv:1307.4924

Radhakrishnan, V., Srinivasan, G.,1982. On the origin of the recently discovered ultra-rapid pulsar CSci, 51, 1096R. https://ui.adsabs.harvard.edu/abs/1982CSci...51.1096R/abstract

Raguzova N.V., Lipunov V.M., 1998 A&A, 340, 85–102 http://adsabs.harvard.edu/abs/1998A%26A...340...85R

Ridnaia, A. et al.,2021. A peculiar hard X-ray counterpart of a Galactic fast radio burst. Nat Astron 5, 372-377. 10.1038/s41550-020-01265-0

Ritchie, B.W. et al.,2010. A VLT/FLAMES survey for massive binaries in Westerlund 1. II. Dynamical constraints on magnetar progenitor masses from the eclipsing binary W13. A&A, 510, A48 10.1051/0004-6361/201014834

Ruiter, A. J. et al., 2019. On the formation of neutron stars via accretion-induced collapse in binaries. MNRAS, 484, 698–711 10.1093/mnras/stz001

Salpeter, E.E., 1964. Accretion of Interstellar Matter by Massive Objects.. ApJ, 140, 796 10.1086/147973

Schenker, K., A.R. King et al.,2002. AE Aquarii: how cataclysmic variables descend from supersoft binaries MNRAS 337, 1105. https://doi.org/10.1046/j.1365-8711.2002.05999.x

Shakura, N.I., Syunyaev, R., 1973. Black holes in binary systems. Observational appearance. A&A , 24, 337-355. http://adsabs.harvard.edu/abs/1973A%26A....24..337S

Shklovsky, I. ,1967. On the Nature of the Source of X-Ray Emission of Sco XR-1. ApJ, 148, 1. http://adsabs.harvard.edu/abs/1967ApJ...148L...1S

Shvartsman, V., 1970. Two generations of pulsars . Radiophysics and Quantum Electronics, 13 (12), 1428-1440 10.1007/BF01032996

Snezhko, L., 1967. On the evolution of close binary systems. PZ, 16, 253S. https://ui.adsabs.harvard.edu/abs/1967PZ.....16..253S/abstract

Suleimanov, V.F., V. Doroshenko, K.Werner, , 2019. Hard X-ray view on intermediate polars in the Gaia era . MNRAS,482, 3, 3622 http://adsabs.harvard.edu/abs/2019MNRAS.482.3622S

Tavani, M. et al., 2020. An X-ray burst from a Magnetar enlightening the mechanism of fast radio bursts. Preprint at https://arxiv.org/abs/2005.12164

Tominaga, N., Blinnikov, S., Nomoto, K., 2013. Supernova Explosions of Super-asymptotic Giant Branch Stars: Multicolor Light Curves of Electron-capture Supernovae. ApJL, 771, L12. 10.1088/2041-8205/771/1/L12

Tutukov, A.V., Yungelson, L., 1973. Evolution of massive close binaries . Nauchnye Informatsii, 27, 70T http://adsabs.harvard.edu/abs/1973NInfo..27...70T

Walker, M.F., 1958. Photoelectric Observations of Nova DQ Herculis (1934). ApJ, 127, 319 http://articles.adsabs.harvard.edu/pdf/1958ApJ...127..319W

Woltjer, L., 1964. X-rays and Type I Supernova Remnants, ApJ, 140, 1309-1312. 10.1086/148028

Zeldovich, Ya.B., 1964. The Fate of a Star and the Evolution of Gravitational Energy Upon Accretion. Soviet Phys. Doklady, 9, 195Z https://ui.adsabs.harvard.edu/abs/1964SPhD....9..195Z/abstract

Zeldovich, Ya.B., Ivanova, L.N., Nadezghin, D.K., 1972. Nonstationary Hydrodynamical Accretion onto a Neutron Star. SvA, 16, 209 http://adsabs.harvard.edu/abs/1972SvA....16..209Z

Zel'dovich,Ya.B., Novikov, I.D., 1966. Relativistic astrophysics.Soviet Physics Uspekhi, V. 8 (4), 522-577 https://www.turpion.org/php/reference.phtml?journal_id=pu&paper_id=2990&volume=8&issue=4&type=xrf

Zeldovich Ya.B., Shakura N.I., 1969. X-Ray Emission Accompanying the Accretion of Gas by a Neutron Star. SvA, 13, 175Z http://adsabs.harvard.edu/abs/1964SPhD....9..195Z